\documentclass[aps,prl,twocolumn,showpacs,superscriptaddress]{revtex4-1}

\usepackage{graphicx}
\usepackage{amssymb}
\usepackage{color}

\newcommand{\rs}{\rm \scriptscriptstyle}

\begin{document}

\title {Scattering resonances and bound states for strongly interacting Rydberg polaritons}
\author{P.  Bienias}
\email[The two authors contributed equally. Corresponding author: ]{przemek@theo3.physik.uni-stuttgart.de}
\affiliation{Institute for Theoretical Physics III, University of Stuttgart, Germany}
\author{S. Choi}
\email[The two authors contributed equally. Corresponding author: ]{przemek@theo3.physik.uni-stuttgart.de}
\affiliation{Physics Department, Harvard University, Cambridge, Massachusetts 02138, USA}
\author{O. Firstenberg}
\affiliation{Physics Department, Harvard University, Cambridge, Massachusetts 02138, USA}
\author{M. F. Maghrebi}
\affiliation{Joint Quantum Institute, National Institute of Standards and Technology, and University of Maryland, College Park, Maryland 20742, USA}
\author{M.~Gullans}
\affiliation{Joint Quantum Institute, National Institute of Standards and Technology, and University of Maryland, College Park, Maryland 20742, USA}
\author{M. D. Lukin}
\affiliation{Physics Department, Harvard University, Cambridge, Massachusetts 02138, USA}
\author{A.V. Gorshkov}
\affiliation{Joint Quantum Institute, National Institute of Standards and Technology, and University of Maryland, College Park, Maryland 20742, USA}
\author{H. P. B\"uchler}
\affiliation{Institute for Theoretical Physics III, University of Stuttgart, Germany}

\date{\today}

\begin{abstract}
We provide a theoretical  framework describing  slow-light polaritons interacting via atomic  Rydberg states. We use a diagrammatic method to analytically derive the scattering properties of two polaritons. We identify  new parameter regimes where polariton-polariton interactions are \textit{repulsive}. Furthermore, in the regime of  \textit{attractive} interactions, we identify multiple two-polariton bound states, calculate their dispersion, and study the resulting scattering resonances. Finally, the two-particle scattering properties allow us to derive the effective low-energy many-body Hamiltonian. 
This theoretical platform is applicable to ongoing experiments.
\end{abstract}


\maketitle

Weak interactions of photons with each other are the basis for many applications of light signals in areas such as optical communication.  However, many other applications in classical and quantum communication, computation, and metrology would greatly benefit from tunable photon-photon interactions. Moreover, photon-photon interactions at the level of individual quanta could pave the way to the realization of exotic strongly correlated photonic states \cite{carusotto13,chang08,Otterbach2013}. A typical approach to achieve strong two-photon interactions relies on confining photons to high-finesse cavities \cite{Fushman2008,Rauschenbeutel1999,Kirchmair2013}. An alternative approach towards this goal has recently emerged using Rydberg slow-light polaritons \cite{Friedler05,pritchard10,Gorshkov2011b,dudin12,parigi12,Peyronel2012b,Firstenberg2013c,petrosyan11}.

The key idea \cite{lukin2001} is to combine electromagnetically induced transparency (EIT) \cite{Fleischhauer2000,Fleischhauer2005} with the strong interaction between Rydberg atoms \cite{Saffman2010}. Both phenomena have been well-studied in the past: it has been demonstrated that photons can be slowed down and stored in atomic gases using EIT \cite{Liu2001,Phillips2001,Julsgaard2004}, while recent experiments on Rydberg atoms have demonstrated the strong interaction and the associated blockade of Rydberg excitations  \cite{Gaetan2009a,Urban2009,Schempp2014,Maxwell2013, Schausz2012,Dudin2012,Heidemann2007}. In the Rydberg-EIT system, a photon entering the atomic gas is converted into a slow-light polariton with a substantial admixture of the Rydberg state. It is the latter admixture that maps the Rydberg-Rydberg interaction onto an effective interaction between slow Rydberg polaritons.  Within this approach, a single-photon source 
\cite{KuzmichScience2012} and switch \cite{Rempe2013} was realized, the photon blockade \cite{Peyronel2012b} and the formation of bound states of Rydberg polaritons \cite{Firstenberg2013c} has been demonstrated, and
atom-photon entanglement was observed 
\cite{KuzmichNature2013}.

%
%

%
\begin{figure}[htp]
\includegraphics[width= 0.95\columnwidth]{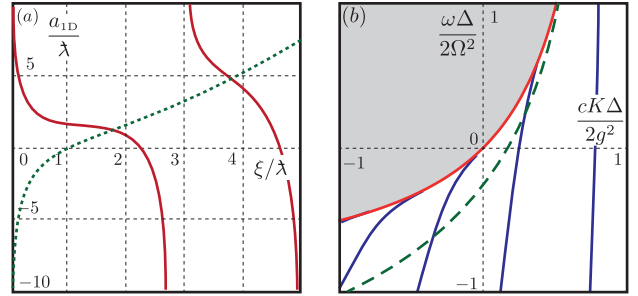}
\caption{
 (a) Low-energy scattering length $a_{\rs 1D}$:
for attractive interactions (solid line), we obtain scattering resonances associated with the appearance of additional bound
states for increasing interaction  strength $\xi/ \lambdabar$;   $\lambdabar = \hbar \sqrt{| \bar{\chi}/\alpha m|}$.  For repulsive interactions (dashed line), we find a single zero crossing.
 (b) Two-polariton spectrum for $\Omega \ll |\Delta| \lesssim g$: For weak interactions $\xi /\lambdabar = 0.5$ (dashed line), we obtain a single bound state below the continuum of scattering states, whereas for strong interactions $\xi /\lambdabar= 5$ (solid lines), we observe the existence of several bound states.
}
\label{fig3}
\end{figure}

In this Letter, we derive the scattering properties and bound-state structure of Rydberg polaritons in one dimension. Our analysis uses  diagrammatic techniques and provides a rigorous theoretical framework for analyzing a variety of problems in Rydberg-polariton systems. This framework allows us to analytically derive the effective interaction potential between two Rydberg polaritons and to identify a regime with a purely repulsive interaction. We derive the low-energy scattering length and find the appearance of resonances, see Fig.~\ref{fig3}; we expect the corresponding tunability of the scattering length to play the role that Feshbach resonances play in ultra-cold atomic gases \cite{chin2010}. Moreover, we identify multiple two-polariton bound states for attractive interactions and determine their dispersion relation. This understanding paves the way for a microscopic derivation of the many-body theory for Rydberg polaritons in the dilute regime.

Before proceeding we note that first steps towards a many-body theory for Rydberg polaritons in one dimension have already been taken \cite{Otterbach2013}. However, a full description of the system -- including the short-range and finite-energy effects relevant to ongoing experiments \cite{Peyronel2012b,Firstenberg2013c} -- is  limited to extended numerical simulations \cite{Gorshkov2011b,MinXiao2014}, which can treat only a  small number of interacting photons.

{\it Model and key results.}---We consider photons inside an atomic ensemble propagating in one-dimension under EIT conditions, 
where the atomic ground state is coupled to a Rydberg state via an intermediate level denotes as $p$-state.
We introduce the electric field operators $\psi^{\dag}_{e}(z)$ and $\psi_{e}(z)$, creating and annihilating a photon at position $z$, respectively.  
If the atomic density is much higher than the photon density, the excitations of atoms are well-described by the bosonic field operators $\psi^{\dag}_{p}(z)$ and $\psi^{\dag}_{s}(z)$. Here, $\psi^{\dag}_{p}(z)$ describes the atomic excitation into the $p$-state, while $\psi^{\dag}_{s}(z)$ generates a Rydberg excitations.
%
We then obtain the non-interacting part of the microscopic Hamiltonian under the rotating-wave approximation in the rotating frame
\begin{equation}
 H_{0} = \hbar \int dz \left(
     \begin{array}{c}
      	    \psi_{e}\\
	    \psi_{p}\\
	    \psi_{s}
     \end{array}\right)^{\dag}
     \left(\begin{array}{ccc}
   - i  c \partial_{z}& g    &  0 \\
         g   &   \Delta &  \Omega\\
        0  &  \Omega & 0
    \end{array} \right)
        \left(
     \begin{array}{c}
      	    \psi_{e}\\
	    \psi_{p}\\
	    \psi_{s}
     \end{array}\right).
     \label{quadraticHamiltonian-1D}
\end{equation}
Here, $g$ denotes the collective coupling of the photons to the matter via the excitation of ground state atoms into the $p$-level, while  $\Omega$ denotes the Rabi frequency of the control field between the $p$-level and the Rydberg state.
Note that the kinetic energy of the photons $-i \hbar \partial_{z}$ only accounts for the deviation from the EIT condition.
We introduced the complex detuning  $\Delta = \delta - i \gamma$, which accounts the detuning $\delta$ of the control field and the decay rate $2 \gamma $ from the $p$-level. Throughout our analysis, we assume $|\delta| \gg \gamma$, thus providing the results in the limit $\gamma =0$. Then,
the inclusion of a finite decay rate $\gamma$ is  obtained by an analytical continuation in $\Delta = \delta - i \gamma$.
%
%
The interaction between the Rydberg levels is described by
\begin{equation}
   H_{\rs rr} = \frac{1}{2} \int dz dz' V(z-z')  \psi_{s}^{\dag}(z) \psi_{s}^{\dag}(z') \psi_{s}(z') \psi_{s}(z).
\end{equation}
In the following, we focus on a van der Waals  interaction $V(r) = C_{6}/r^6$. The microscopic Hamiltonian $H_{0}+H_{\rs rr}$ describes three bosonic fields with a non-interacting quadratic part and a quartic interaction.
Such systems have been extensively studied in the past using diagrammatic methods, \emph{cf.}~Ref.~\cite{agd}. However, it is important to stress that  the quadratic
Hamiltonian exhibits a rather unconventional form, as the only dynamics is given by the light velocity of the photon. It is this property, together with the conservation of total energy $\hbar \omega$ and total momentum $\hbar K$, that is crucial to our analysis and gives rise to novel phenomena.

In order to understand the many-body properties of the system, we will analyze the scattering 
properties and bound-state structure for two  polaritons.  The main idea is to derive the
scattering length $a_{\rs 1D}$, which in turn allows for the description of the many-body theory in terms 
of a pseudo-potential. 
%
This approach is in analogy to cold atomic gases, where it is extremely successful \cite{chin2010}. 

The description of two polaritons requires, in the most general case, a  nine-component, 
two-particle wavefunctions $\psi_{\alpha \beta}(z,z')$ with $\alpha, \beta \in \{e, p, s \}$, 
which denotes the amplitude of finding particles in states $\alpha$ and $\beta$ at $z$ and $z'$, respectively \cite{Gorshkov2011b}. 
To utilize the conservation of energy and momentum, we rewrite $\psi_{\alpha \beta}$ in the center-of-mass $R = (z+z')/2$ and relative $r = z - z'$ coordinates and parametrize it in terms of temporal and spatial Fourier components ($\omega, K$), leaving $r$ the only degree of freedom. Using the diagrammatic techniques, we find that the quantum dynamics of the two polaritons is well captured by a Schr\"odinger-like equation for a single component
\begin{equation}
   \hbar \bar{\omega} \psi(r) = \left[ -\frac{\hbar^2}{m} \partial_{r}^2 +\alpha V_{\rs eff}(r) \right] \psi(r),
   \label{effectiveequation}   
\end{equation}
with the effective potential and polariton mass
\begin{equation}
  V_{\rs eff} (r) = \frac{V(r)}{1- \bar{\chi}(\omega) V(r)},  \hspace{20pt} m =\hbar \frac{(g^2+\Omega^2)^3}{ 2 c^2 g^2 \Delta \Omega^2}.
  \label{effectiveinteraction}
\end{equation}
Here, $\hbar \bar{\omega}(K,\omega)$ is the energy due to the relative motion, and the dimensionless parameter $\alpha(K,\omega)$ can be interpreted as the overlap of the polaritons with the Rydberg state. 
The effective interaction potential  is renormalized as the interaction shifts the two 
Rydberg states out of resonance, when two polaritons approach each other. This behavior is captured 
by a single parameter  $\bar{\chi}(\omega)$.
%
%
The wavefunction $\psi(r)$ is related to the two-body wavefunctions via $\psi(r) =\psi_{ss}(r) [1-\bar{\chi} V(r)]$, where the amplitude $\psi_{ss}(r)$ to find two Rydberg states 
 vanishes at distances shorter than
the blockade radius $\xi =(|C_{6} \bar{\chi}|)^{1/6}$. In addition, the wavefunction $\psi(r)$ is proportional to the electric field amplitude $\psi_{ee}(r)$.
Equation~(\ref{effectiveequation}) is valid in several  relevant regimes, including the low-momentum and low-energy regime, and the far-detuned regime as discussed below.  For vanishing  $\omega =0$,  Eq.~(\ref{effectiveequation}) reproduces the steady state results obtained in Refs~\cite{Gorshkov2011b,Firstenberg2013c}.

We start with the  \emph{low-momentum and low-energy regime}, which allows us to analytically derive the low-energy scattering length $a_{\rs 1D}$.  The characteristic energy scale is given by $ \omega_{c} =  \min\{|\Delta|, 2 \Omega^2/|\Delta|\}$
and the corresponding characteristic momentum  is $ q_{c} = \omega_{c}/v_{g}$ with $v_{g} = \Omega^2/(\Omega^2+g^2) c$ the slow light velocity.
At low energies $|\omega| \ll \omega_{c}$ and momenta $|K| \ll q_{c}$, the expressions for 
$\bar{\omega}$ and $\alpha$ are in  leading order
\begin{equation}
   \bar{\omega}= \omega - v_{g} K, \hspace{20 pt} \alpha = \frac{g^4}{(g^2+\Omega^2)^2} .
\end{equation}
In this limit, Eq.~(\ref{effectiveequation}) provides the intuitive result: $\bar{\omega}$ is the difference between the total energy $\omega$ and the kinetic energy $v_g K$ of the center-of-mass motion, and $\alpha$ is the square of the probability to find a 
polariton in the Rydberg state.  Furthermore, the amplitude for the electric field takes the form $\psi_{ee}= \Omega^2/g^2 \psi$.
At the same time, $\bar{\chi}$ reduces to
\begin{equation}
    \hbar \bar{\chi} = \frac{\Delta}{2 \Omega^2} - \frac{1}{2 \Delta},
\end{equation}
which exhibits a zero crossing for $\Omega=\pm \Delta$ with an associated sign change.
Therefore, it is possible to realize effectively \emph{repulsive} polariton-polariton interactions for $\Omega >|\Delta|$ with $C_{6} \delta >0$ (the latter condition avoids a singularity in $V_{\rs eff}$). This surprising behaviour is  related to the modification  of frequency-dependant part of susceptibility of proximal polaritons.
It
is in contrast to  the far off-resonant regime ($\Omega \ll |\Delta|$),  where the combination of polariton mass and effective interaction always leads to an effective attraction \cite{Firstenberg2013c}. 
In this regime, we observe a transition from a negative to a positive 1D scattering-length $a_{\rs 1D}$ for increasing interactions, see Fig.~\ref{fig3}(a).
The interaction strength is conveniently expressed by the dimensionless parameter $\xi/\lambdabar$ with  the blockade radius $\xi =(| C_{6} \bar{\chi}|)^{1/6} $ and $\lambdabar =\sqrt{ | \hbar^2\bar{\chi} / (\alpha m)|}$ the de Broglie wavelength associated with the depth/height of the effective potential. Then, we obtain the asymptotic behavior
$a_{\rs 1D} =- (3/\pi)  \lambdabar^2 /\xi$, 
valid for weak interactions with $\xi/\lambdabar \ll 1$,
where the interaction potential can be replaced by a $\delta$-function. Note that for $\gamma =0$, the scattering length is negative, while  for a finite decay rate $\gamma >0$,  the  analytical continuation of the scattering length reduces to $a_{\rs 1D} =(3 /\pi) (- \bar{\chi}^5/ C_{6})^{1/6}(\hbar^2/\alpha m) $ and gives rise to an imaginary contribution  accounting for losses from the $p$-level during the collision. For increasing interactions, eventually a zero crossing of $a_{\rs 1D}$ appears, and we obtain the positive scattering length  $a_{\rs 1D} \approx 0.7 \left( \alpha m C_{6} / \hbar^2 \right)^{1/4}$, where the full tail of the van der Waals interaction dominates. 

In the \emph{attractive} regime $\Omega < |\Delta|$ with $C_{6} \delta <0$, the system generally gives rise to bound states.
Note that bound states can be identified by negative values of $\bar{\omega}$.  
For weak interaction $\xi< \lambdabar$, a single bound state is present, and we recover the expression for the scattering length $a_{\rs 1D} = (3/\pi) \lambdabar^2 /\xi$, which is now positive. 
For increasing interactions $\xi> \lambdabar$, additional bound states will appear.    Each additional bound state is associated with a  resonance in the scattering length in analogy to Feshbach resonances in cold atomic gases \cite{chin2010}. The exact determination of the scattering length $a_{\rs 1D}$
requires the full treatment of the effective interaction potential $V_{\rs eff}(r)$; the latter is easily achieved numerically, see Fig.~\ref{fig3}(a).
It clearly demonstrates that  we can tune the scattering length to arbitrary values by controlling the single parameter $\xi /\lambdabar$,  which defines  the strength of the interaction potential.

In general, the bound states will violate the condition of low energy and are thus more appropriately studied next in the \emph{far-detuned regime} with $\Omega \ll |\Delta|$,  which is valid for all momenta $K $ with the weak constraint on the energy $|\omega| \ll  |\Delta|$.
In this regime, 
we obtain $ \hbar \: \bar{\chi}(\omega) = (\omega + 2 \Omega^2/\Delta)^{-1}$, and the blockade radius reduces to $\xi = (|C_{6} \Delta/2\Omega^2|)^{1/6}$,
while the analytic but lengthy expressions for  $\bar{\omega}$ and $\alpha$ are presented in the supplementary information.
In the experimentally most interesting regime of slow light  $g \gg \Omega$ with $g\gtrsim |\Delta|$, we find
\begin{equation}
\alpha  =  \frac{ 1 - \frac{c K \Delta}{2 g^2}}{\left( 1 + \frac{ \omega \Delta}{2 \Omega^2}\right)^2},
\label{alphadetuned}
\end{equation}
while the expression for the energy $\bar{\omega}$ takes the form
\begin{equation}
 \frac{  \bar{\omega} \Delta}{ 2 \Omega^2} = \frac{\frac{\omega \Delta}{2 \Omega^2 }}{1+ \frac{ \omega \Delta}{2 \Omega^2}}   - \frac{1+ 2 \frac{\omega \Delta}{2 \Omega^2}}{1+ \frac{\omega \Delta}{2 \Omega^2}}\:  \frac{c K \Delta}{2 g^2} + \left(\frac{  c K \Delta}{2 g^2}\right)^2  .
   \label{onshell}
\end{equation}
%
Finally, the relation to the electric field amplitude $\psi_{ee} $  is again closely related to the wavefunction $\psi$ via $(g^2- c K\Delta/2 )\psi_{ee}= (\Omega^2+ \omega \Delta/2) \psi$.
It is important to stress that, in this limit, the diagrammatic result agrees with the approach utilizing adiabatic elimination, see supplementary information. 

%
%

The effective equation Eq.~(\ref{effectiveequation}) allows us to derive the bound states and their group velocity in addition to the scattering states. The spectrum  is shown in Fig.~\ref{fig3}(b): it exhibits a continuum of scattering states as well as bound states. Note that the interaction potential as well as $\bar{\omega}$ depend on the energy $\omega$, and therefore the bound-state energies have to be determined self-consistently. 
The dimensionless parameter measuring the strength of the interaction reduces to
$\xi/\lambdabar = \xi g^2/| \Delta| c$ and is related to the resonant optical depth per blockade radius 
 $\kappa_{\xi} = 2 \xi |\Delta |/ \lambdabar \gamma$. 
Then for weak interactions  $\xi/\lambdabar  < 1$, we  recover a single bound state, which is well described by replacing the effective interaction potential by a $\delta$-function.   For increasing interaction strength $\xi/\lambdabar  >1$, we observe the appearance of additional bound states. Then, the exact bound state energy requires the numerical treatment of the full effective interaction potential Eq.~(\ref{effectiveinteraction}). The result is shown in Fig.~\ref{fig3}(b) for two different interaction strengths.


{\it Derivation and limitations.}---Next, we present the microscopic derivation  of Eq.~(\ref{effectiveequation}) and discuss its limitations. We start by diagonalizing the non-interacting part of the Hamiltonian in momentum space obtaining the dispersion relations of three polariton modes, 
$H_{0}= \sum_{q,\alpha \in {0, \pm 1}} \epsilon_{\alpha q}
 \bar{\psi}^{\dag}_{\alpha q} \tilde{ \psi}_{\alpha q}$.
Here, $\alpha \in {\pm 1}$ account for the two bright polariton modes, while $\alpha = 0$ denotes the dark  polariton.
The new field operators take the form $ \tilde{\psi}_{\alpha q} = \sum_{\beta \in \{e,p,s\}} U_{\alpha}^{\beta}(q) \psi_{\beta q} $ with $\alpha \in \{0, \pm 1\} $, and the inverse $\bar{U} \equiv U^{-1}$ provides  $\bar{\psi}^{\dag}_{\alpha q} = \sum_{\beta \in \{e,p,s\}} \bar{U}_{\beta}^{\alpha}(q) \psi^{\dag}_{\beta q} $.

\begin{figure}[htp]
\includegraphics[width= 1\columnwidth]{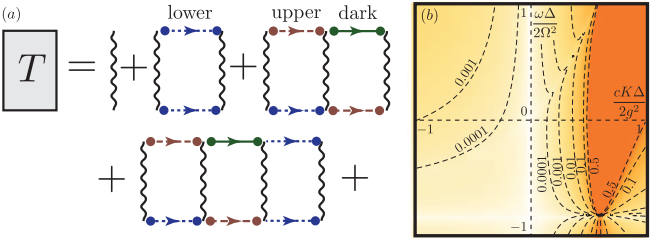}
\caption{(a) Illustration of a  few ladder diagrams: the interaction $V(r)$ is denoted by a wavy line, while the straight lines with an arrow are  Green's functions for the three polariton modes $1/(\hbar \omega - \epsilon_{\alpha} + i \eta)$, and the dots mark the overlap factors $U^{s}_{\alpha}$  and $\bar{U}_{s}^{\alpha}$ of the polariton with the Rydberg state.  (b) Parameter $\zeta(K,\omega)$ measuring the influence of the second pole for $g \gg \Omega$ and $\Omega/\Delta = 0.5$. In the low energy/momentum limit, the second pole can be safely neglected, however its influence strongly increases for $K c \Delta /2 g^2 \sim 1$.
}
\label{fig2} 
\end{figure}

The two-polariton scattering properties  are well accounted for by the $T$-matrix.
As the interaction acts only between the
two Rydberg states, it is sufficient to study the $T$-matrix for the Rydberg states alone, denoted as $T_{k k'}(K,\omega)$.
Here, $k$ is the relative momentum of the two incoming  polaritons and $k'$ the relative momentum of the
outgoing polaritons.  For two polaritons, the $T$-matrix is
expressed as a resummation of all ladder diagrams, see Fig.~\ref{fig2}(a), 
and gives rise to the integral equation  \cite{agd}
\begin{displaymath}
T_{kk'}(K,\omega) = V(k-k') + \int \frac{d q}{2\pi}  V(k-q) \chi_{q}(K,\omega) T_{qk'}(K,\omega).
\end{displaymath}
The full pair propagator of two polaritons and its overlap with the Rydberg state takes the form
\begin{equation}
  \chi_{q}\left(K,\omega\right) = \sum_{\alpha,\beta \in \{0,\pm 1\}}  \frac{\bar{U}^{\alpha}_{s }(p)  U_{\alpha}^{s}(p)   \bar{U}^{\beta}_{s }(p')U_{\beta}^{s}(p') }{\hbar \omega - \epsilon_{\alpha}(p) - \epsilon_{\beta}(p') + i \eta},
\end{equation}
with $q=(p-p')/2$ and $K=p+p'$. It is a special property of our polariton Hamiltonian that
the pair propagation reduces to three terms,
\begin{equation}
\chi_q = \bar{\chi}+ \frac{\alpha}{\hbar \bar{\omega}-\hbar^2 q^2/m+i \eta} + \frac{\alpha_{\rs B}}{\hbar \bar{\omega}_{\rs B} -\hbar^2 q^2/m + i \eta}
 \label{chiexpansion}.
\end{equation}
Here, $\bar{\chi}(\omega)$ accounts for the saturation of the pair propagation at large momenta $q\rightarrow \pm \infty$
and takes the  form
\begin{equation}
\bar{\chi}(\omega) =\frac{1}{\hbar} \frac{ \Delta-\frac{\omega}{2} - \frac{\Omega^2}{\Delta - \omega}}{ \omega\left(\Delta - \frac{\omega}{2}\right) + 2 \Omega^2}.
\label{chi}
\end{equation}
The second term in Eq.~(\ref{chiexpansion}) is  the pole structure for the propagation of the two incoming
polaritons. This term reduces to the propagator of a single massive particle, where  $\alpha$ and $\bar{\omega}$
depend on the center-of-mass momentum $K$ and total energy $\omega$. The latter defines  the relative 
momentum $k = \pm \sqrt{m \bar{\omega}/\hbar}$ of the incoming scattering states. Finally, the last term 
accounts for a second pole, describing the phenomenon of resonant scattering of two incoming polaritons into 
a different outgoing channel, \emph{e.g.}, the conversion of two dark  polaritons into an upper and a lower bright polariton.
The influence of the second pole is measured by the dimensionless parameter
$\zeta(K,\omega)=\sqrt{|(\bar{\omega}  \alpha_{\rs B}^2)/(\bar{\omega}_{\rs B}\alpha^2)|} $. Especially, 
$\zeta(K, \omega) $ is strongly suppressed in the two regimes discussed above; an illustration demonstrating
the strong suppression is shown Fig.~\ref{fig2}(b), while the analytical expressions are provided in the supplement materials.
In these cases the second pole can be dropped in leading order in the small parameter $\zeta \ll 1$.
The saturation $\bar{\chi}$ can
be eliminated by introducing the effective interaction potential $V_{\rs eff}(r)$ given in Eq.~(\ref{effectiveinteraction}).
Then, the equation for the $T$-matrix reduces to
\begin{displaymath}
T_{k k'} = V_{\rs eff}(k-k') + \int \frac{d q}{2 \pi}  V_{\rs eff}(k-q) \: \frac{\alpha}{ \hbar \bar{\omega}-\hbar^2 q^2/m} \:T_{qk'}.
\end{displaymath}
Consequently the $T$-matrix describes a system of a single massive particle in the 
effective interaction potential $V_{\rs eff}$ with the relative coordinate as the degree of freedom and
is fully described by the Schr\"odinger equation in Eq.~(\ref{effectiveequation}).
The relation  $\psi(r) = \psi_{ss}(r) [1-\bar{\chi} V(r)]$ follows from the relation between the $T$-matrix and the
scattering wave function $\psi_{ss}(r)  V(r)= \int  d k'    e^{i r k' }  T_{k k'}/(2 \pi) = \psi(r) V_{\rs eff}(r)$.

{\it Discussion and Outlook.}---The full understanding of the scattering properties allows us  to derive the \emph{low energy many-body Hamiltonian for Rydberg polaritons}. Here, the fundamental assumption is that each scattering process of the polaritons is independent of each other. This condition is  satisfied in the dilute regime $n_{d} \: r_{0} \ll 1$, where the density $n_{d}$ of Rydberg polaritons is low compared to the range $r_{0}$ of the interaction potential. The latter
is determined either by the blockade radius or the van der Waals length, \emph{i.e.},
$r_{0} = \max\{ \xi, (|\alpha m C_{6}|/\hbar^2)^{1/4}\}$.
Then, the interaction is fully determined by the scattering length  $a_{\rs 1D}$ via the one-dimensional pseudo-potential
$V_{\rs 1D} = -2 \hbar^2 /m a_{\rs 1D} \delta(r) $  \cite{Olshanii1998},
and the many-body theory reduces to the Hamiltonian
\begin{displaymath}
 H \! = \!  \int \! dx \!  \left[ \psi^{\dag}_{d} \left( - i \hbar v_{g}  \partial_{z}- \frac{\hbar^2}{2 m} \partial_{z}^2 \right) \psi_{d}  - \frac{2 \hbar^2}{m  a_{\rs 1D}} \psi^{\dag}_{d}\psi^{\dag}_{d}\psi_{d}\psi_{d} \right] , 
\end{displaymath}
with $\psi_{d}^{\dag}$ ($\psi_{d}$) denoting the bosonic field operator creating (annihilating) a Rydberg polariton.
Here we can control the scattering length $a_{\rs 1D}$ by the strength of the interactions, see Fig.~\ref{fig3}(a). We can therefore study continuously  the crossover from a Lieb-Liniger gas at $a_{\rs 1D} < 0$ to the Super-Tonks-Girardeaux gas at  $a_{\rs 1D} > 0$ by tuning the parameters through a zero crossing of the scattering length \cite{Astrakharchik2005,Chen2010,Girardeau2010}. In contrast to cold atomic gases \cite{Haller2009,Sala2012},  the zero crossing of the scattering length is not associated with losses in the system.
%

Finally, we point out that a complementary derivation of an effective low-energy theory can also be achieved at high densities, if the interaction is dominated by the purely repulsive part of the van der Waals interaction, as proposed in Ref.~\cite{Otterbach2013}.  
This theory is of interested in the parameter regime with   $ 1/(|\alpha m C_{6}|/\hbar^2)^{1/4} < n_{\rs 1D} < q_{c}, 1/|\xi|$; note that here we provide a microscopic derivation for the short distance behavior.  
We have demonstrated that this regime is most interesting to study when $\Delta \approx \pm  \Omega$ and $C_{6} \delta > 0$, where $\bar{\chi}$ is strongly suppressed,  and the effective interaction reduces to the pure van der Waals repulsion $V_{\rs eff }(r) = C_{6}/r^6$. 
This scenario allows one to observe the crossover into a regime  where crystalline correlations dominate the ground state.

{\it Experimental implications.}---The microscopic analysis presented here has several implications for experiments. First, the existence of a parameter regime with a purely repulsive interaction will give rise to photon anti-bunching for the two-photon correlations in an experimental setup similar to that of Ref. \cite{Firstenberg2013c}. The experimental requirements are strong Rabi frequency $\Omega\gtrsim| \Delta|$, $q_{c} \xi < 1$  for the low momentum regime, and $\gamma \ll |\delta|$ to distinguish the repulsion from losses.
 In turn, the analysis of the bound-state structure allows for the determination of the group velocity. As can be seen in Fig.~\ref{fig3}(b), the group velocity of the bound states is larger than the slow light velocity, and the bound states will travel ahead of the continuum. This will allow one to spatially separate the bound photon pairs in a pulsed experiment. Finally, the scattering length defines the phase shift two polaritons pick up during a collision; it has been proposed to use such collisions to realize photonic two-qubit gates \cite{Friedler05,Gorshkov2011b}. Here, the predicted zero-crossing of the scattering length corresponds to the optimal $\pi$-phase shift. A direct measurement of these resonances is possible in a setup with  frequency difference $\Delta \omega$ and spatially-resolved detection of the polaritons inside the medium. Therein, the correlation function in the relative coordinate will oscillate with a wavevector $\Delta k = \Delta \omega/ v_{g}$. The maxima of these oscillations will shift for increasing scattering length by a phase $\phi$ via  $\cot(\phi) =  - a_{\rs 1D} \Delta k $. The details of these observations depend on the experimental setup and on the precise boundary conditions but can be efficiently addressed within the presented framework.

{\bf Acknowledgements:} We thanks T. Pohl, M. Fleischhauer, P. Strack, and M. Hafezi for discussions. 
 We acknowledge support by the Center for Integrated Quantum Science and Technology (IQST), the Deutsche Forschungsgemeinschaft (DFG) within SFB TRR 21, the EU  Marie Curie ITN COHERENCE, CUA, AFOSR MURI, DARPA QUINESS, Packard Foundation,  and  the National Science Foundation.
A.G and M.M  thank  the JQI and the PFC at the JQI for funding.  HB, ML, AG acknowledge hosptiality of the KITP.

\section{ Supplement Material}

{\bf Influence and strength of the second pole:}
In the following, we estimate the relevance of the second pole in Eq.~(10) in the main text, characterized by  $\alpha_{\rs B}$ and
$\bar{\omega}_{\rs B}$, which gives rise to the resonant scattering into a different outgoing channel.
%
%
First, we concentrate on the low momentum and energy regime. The analytical expressions for
 $\alpha$ and $\bar{\omega}$ describing the first pole are given by Eq.~(5) in the main text.
 In turn, the parameters for the second pole derived by the diagrammatic method take the form
 %
%
%
\begin{eqnarray}
\alpha_{\rs B} &=& -\frac{ (\omega-c K )^2 \Omega^6}{4 \Delta^2  (g^2+\Omega^2)^3},\\
\bar{\omega}_{\rs B} &=& \frac{4 \Omega^2 g^4}{(g^2+\Omega^2)^3}\frac{\Delta^2 }{\omega- c K }.
\end{eqnarray}
Note that the weight $\alpha_{\rs B}$ of the pole vanishes  quadratically in the low energy $|\omega | \ll \omega_{c}$
and momentum $| K|  \ll q_{c}$ limit, and can therefore be safely dropped.

Next, we analyse the influence of the second pole in the  regime of far-detuned Rydberg polaritons with 
$|\omega|,\Omega\ll |\Delta|,g$. The diagrammatic approach provides the analytic expressions
%
%
%
\begin{eqnarray}
\alpha_{\rs B} &=&-  \frac{ \Omega ^6 (1+\frac{ c K}{2 \Delta}) (\omega -c K)^2}{4 \Delta^2  (g^2+\Omega^2)^3 \left(1- \frac{ c K \Delta}{2 g^2}\right)^2},\\
\bar{\omega}_{\rs B}&=&- \left(1+\frac{c K}{2 \Delta}\right)^2 \left(1-\frac{ c K \Delta}{2 g^2}\right)  \frac{4 \Omega^2 g^4}{(g^2+\Omega^2)^3}\frac{\Delta^2 }{c K-\omega }. \nonumber
\end{eqnarray}
We find that, in the regime  $ c K \delta/2g^2 < 1$, the dimensionless parameters
$\zeta(K,\omega)$ is strongly  suppressed by the factor $(\Omega/\Delta)^6$. However, it is important to stress that the strength of the second pole diverges in a narrow parameter regime around  $ c K \delta/2 g^2  \approx 1$.

{\bf Adiabatic elimination:} In the following, we compare our diagrammatic approach with the previously successfully-applied
study of the two-particle equation for the wave function in the regime $\omega =0$, where the $p$-level has  sometimes been adiabatically eliminated \cite{Gorshkov2011b,Peyronel2012b,Firstenberg2013c}. Furthermore, we  present the natural extension of adiabatic elimination for finite frequencies.
Then, the two-particle wave function contains four components: $\psi_{ee}$ describes the amplitude for two photons, $\psi_{ss}$ the amplitude for two Rydberg atoms,  and $\psi_{es\pm}$ the amplitude for one photon and one Rydberg atom with even (odd) symmetry.
The Schr\"odinger equation reduces to (see Refs.~\cite{Gorshkov2011b,Peyronel2012b,Firstenberg2013c} for more details)
\begin{eqnarray}
 \omega \psi_{ee} & =& - i c \partial_{R}  \psi_{ee} - \frac{2 g^2}{\Delta} \psi_{ee} - \frac{2 g \Omega}{\Delta} \psi_{es+}, \label{EE}\\
 \omega \psi_{es+} &=& -\frac{ i c}{2} \partial_{R} \psi_{es+} - i c \partial_{r} \psi_{es-}  \nonumber\\
 & &\hspace{10pt} - \frac{g^2+ \Omega^2}{\Delta} \psi_{es+} -\frac{g \Omega}{\Delta} \left(\psi_{ee}+\psi_{ss}\right), \label{ESp}\\
  \omega \psi_{es-}& =& -\frac{i c}{2} \partial_{R}\psi_{es-} - i c \partial_{r} \psi_{es+}-\frac{g^2 + \Omega^2}{\Delta}  \psi_{es-}, \label{ESm}\\
 \omega \psi_{ss}&  = &-\frac{2 \Omega^2}{\Delta} \psi_{ss} - \frac{2 g \Omega}{\Delta} \psi_{es+} +\frac{ V(r)}{\hbar}  \psi_{ss} \label{SS},
\end{eqnarray}
where  $r$ denotes the relative coordinate and $R$ the center-of-mass coordinate. For the translational
invariant system, the latter coordinate is expressed in Fourier space with $K$ the total momentum.
We can solve the first, third and fourth equations for $\psi_{ee}$, $\psi_{es-}$, and $\psi_{ss}$, respectively,
%
Inserting these expressions into Eq.~(\ref{ESp}), we obtain a single differential equation involving only
the wave function $\psi_{es+}$,
\begin{equation}
 \hbar \bar{\omega} \psi_{es+} = - \frac{\hbar^2}{m}\partial_{r}^2 \psi_{es+}  + \alpha V_{\rs eff}(r) \psi_{es+}.
\end{equation}
This equation takes exactly the form of Eq.~(\ref{effectiveequation}) in the main text with the identification
$\psi_{es+}  \sim \psi$. The expressions for $\alpha$ and $\bar{\omega}$ within the adiabatic elimination reduce
to
\begin{eqnarray}
\frac{\alpha m}{\hbar^2} & =& \frac{g^2 \Omega^2}{c^2 \hbar \Delta^2} \frac{  2 (\omega + \frac{g^2+\Omega^2}{\Delta}) - c K}{\left( \omega + 2\frac{ \Omega^2}{\Delta}\right)^2}, \\    \bar{\chi} & = &\frac{1}{\hbar} \frac{1}{\omega + 2 \Omega^2/\Delta}, \nonumber\\
 \frac{\bar{\omega} m}{\hbar}& =& \left[  c K -2 (\omega + \frac{\Omega^2+g^2}{\Delta})\right]^2 \\
 & &\hspace{10pt } \times \frac{2 \omega \frac{\Omega^2 +g^2}{\Delta}+\omega (\omega- c K) -\frac{2 \Omega^2}{ \Delta} c K }{4 c^2 \left(\omega + \frac{2 \Omega^2}{ \Delta}\right) \left(\omega -c K+\frac{2 g^2}{\Delta}\right)}  .
 \nonumber
\end{eqnarray}
These expressions fully agree with the result dervied within the diagrammatic approach in the limit of large detuning $\Omega \ll |\Delta|$ and energies $|\omega| \ll |\Delta| $.
In the physically interesting situation of Rydberg polaritons with $g\gtrsim |\Delta|$, we finally obtain the expressions (\ref{alphadetuned})
and (\ref{onshell}) in the main text.

\vspace{-15pt}

\bibliography{bib}


\end{document}